\documentclass[     ,final              ]{aipproc}%
\usepackage{amssymb}
\usepackage{amsmath}
\usepackage{amsfonts}
\usepackage{graphicx}%
\input{aipcheck}
\layoutstyle{6x9}
\begin{document}


\title{ Pressure of the $O(N)$ Model in $1+1$ Dimensions}
\author{Francesco Giacosa}{address={
    Institut f\"{u}r Theoretische Physik, Johann Wolfgang
    Goethe-Universit\"{a}t,\\ Max-von-Laue-Str.~1, D-60438 Frankfurt, Germany
  }
}
\author{Stefano Lottini}{address={
    Institut f\"{u}r Theoretische Physik, Johann Wolfgang
    Goethe-Universit\"{a}t,\\ Max-von-Laue-Str.~1, D-60438 Frankfurt, Germany
  }
}
\author{Elina Seel}{address={
    Institut f\"{u}r Theoretische Physik, Johann Wolfgang
    Goethe-Universit\"{a}t,\\ Max-von-Laue-Str.~1, D-60438 Frankfurt, Germany
  }
}
\author{Dominik Smith}{address={
    Institut f\"{u}r Theoretische Physik, Johann Wolfgang
    Goethe-Universit\"{a}t,\\ Max-von-Laue-Str.~1, D-60438 Frankfurt, Germany
  }
  ,altaddress={
		Fakult\"at f\"ur Physik,
		Universit\"at Bielefeld, \\
		Universit\"atsstr.~25,
		D-33615 Bielefeld, Germany
	}
}

\begin{abstract}
The $O(N)$ model in $1+1$ dimensions presents some features in common with
Yang-Mills theories: asymptotic freedom, trace anomaly, non-petrurbative
generation of a mass gap. An analytical approach to determine the
termodynamical properties of the $O(3)$ model is presented and compared to
lattice results. Here the focus is on the pressure: it is shown how to derive
the pressure in the CJT formalism at the one-loop level by making use of the
auxiliary field method. Then, the pressure is compared to lattice results.

\end{abstract}
\maketitle

%%%%%%%%%%%%%%%%%%%%%%%%%%%%%%%%%%%%%%%%%%%%
%% MAINMATTER
%%%%%%%%%%%%%%%%%%%%%%%%%%%%%%%%%%%%%%%%%%%%

\section{Introduction}

The $O(N)$ model, in $1+1$ dimensions and at nonzero temperature $T$, is
defined by the following generating functional:
\begin{equation}
\ Z_{O(N)}=\mathcal{N}\int\mathcal{D}\Phi\delta\left(  \Phi^{2}-\frac{N}%
{g^{2}}\right)  \exp\left[  - \overset{\ }{\int_{0}^{\beta}}d\tau\int_{-\infty
}^{\ \infty}dx\mathcal{L}_{0}\right]  \text{ ,} \label{z}%
\end{equation}
whereas $g$ is the dimensionless coupling constant, $\mathcal{N}$ is a
normalization constant, and $\mathcal{L}_{0}$ is a simple free (Euclidean) Lagrangian
$\mathcal{L}_{0}=\dfrac{1}{2}\partial_{\mu}\Phi^{t}\partial_{\mu}\Phi$ , with
$\Phi^{2}=\Phi^{t}\Phi$.
$\Phi$ consists of $N$ scalar real fields, which for future convenience we 
denote as $\Phi^{t}=\left(  \sigma,\pi_{1},\ldots,\pi_{N-1}\right)$.
The fields are constrained by the condition $\Phi
^{2}=N/g^{2}$ which is incorporated by the delta function in Eq.~(\ref{z}).
This nonlinear constraint enforces the thermodynamics of the model on an $N-1$
dimensional hypersphere and induces the interactions between the fields. This
model is interesting because it shares some properties with Yang-Mills
theories in 4 dimensions, see the review paper \cite{novikov} and refs.~therein:

(i) The coupling constant $g$ is dimensionless and the the theory is
renormalizable. It turns out that it has a negative $\beta$ function, thus
showing asymptotic freedom.

(ii) The model is conformal invariant at the classical level, but, just as
Yang-Mills theories in four dimensions, at the quantum level an energy scale
emerges due to quantum corrections and a non-perturbative mass-gap emerges
(trace anomaly) \cite{wolff}: although the Lagrangian $\mathcal{L}_{0}$ in Eq.
(\ref{z}) describes $N$ massless fields, a nonzero mass $m=\mu\exp(-2\pi
/g^{2})$ is generated dynamically due to the interactions, where $\mu$ is the
renormalization parameter. Since the mass is non-analytic in $g$, it would
vanish in perturbation theory about $g=0$.

(iii) For $N=3$ the model has instantons and, at finite temperature, calorons
\cite{bruckmann}. It is then possible to study in a simplified framework what
is the role of these nonperturbative field configurations in
thermodynamics (for the related case in YM theories see for instance 
refs.~\cite{ralf1,pisarski} and refs.~therein).

The aim of this work is to present an analytical study of the nonlinear $O(N)$
model in $1+1$ dimensions at nonzero temperature. We compute the pressure by
employing the CJT formalism \cite{cjt} and using the auxiliary field method in
the one-loop approximation (for other thermodynamical quantities and the
two-loop calculation and technical details see ref.~\cite{our}). For the case
$N=3$ we compare the pressure with the results of a \textquotedblleft
first-principles\textquotedblright\ Monte Carlo lattice calculation using the
integral method. At the one-loop level, a good
agreement is found for small temperature, but the analytic result for the
pressure is too large when $T$ increases. When going to the two-loop
approximation, a considerable improvement in the high $T$ domain is obtained,
although the low-$T$ part is slightly worsened.

\section{Analytic approach and results}

Following ref.~\cite{seel} we rewrite the partition function in Eq.~(\ref{z})
by using the mathematically well-defined (i.e.~convergent) form of the
$\delta$-function%
\[
\delta\left(  \Phi^{2}-\dfrac{N}{g^{2}}\right)  =\lim_{\varepsilon
\rightarrow\text{ }0^{+}}N\int\mathcal{D}\alpha e^{\left\{  -\ \overset
{\ }{\int_{0}^{\beta}}d\tau\int_{-\infty}^{\ \infty}dx\left[  \frac{i\alpha
}{2}\left(  \Phi^{2}-\frac{N}{g^{2}}\right)  +\frac{\varepsilon\alpha^{2}}%
{2}\right]  \right\}  }\text{ }%
\]
thus obtaining $Z_{O(N)}=\lim_{\varepsilon\rightarrow\text{ }0^{+}}%
\mathcal{N}\int\mathcal{D}\alpha\mathcal{D}\Phi\exp[-\overset{\ }{\int
_{0}^{\beta}}d\tau\int_{-\infty}^{\ \infty}dx\mathcal{L]}$, where%
\begin{equation}
\mathcal{L}=\dfrac{1}{2}\partial_{\mu}\Phi^{t}\partial_{\mu}\Phi+U(\Phi
,\alpha)\text{ with }U(\Phi,\alpha)=\dfrac{i}{2}\alpha\left(\Phi^{2}-\dfrac
{N}{g^{2}}\right)+\dfrac{\varepsilon}{2}\alpha^{2}\text{ .}%
\end{equation}
The quantity $\alpha$ is an auxiliary field serving as a Lagrange multiplier.
Upon shifting the fields $\sigma$ and $\alpha$ by their condensates,
$\sigma\rightarrow\phi+\sigma$ and $\alpha\rightarrow\ \alpha_{0}+\alpha$, a
bilinear mixing term of the type $-i\alpha\sigma\phi$ arises. It can be
eliminated by a further shift of the $\alpha$ field, $\alpha\longrightarrow
\alpha-4i\phi\sigma/N\varepsilon.$ In this way non-diagonal propagators are
avoided.

Within the so-called CJT formalism \cite{cjt} the standard expression for the
effective potential is given by%
\[
V_\mathrm{eff}=U(\phi,\alpha_{0})+\sum_{i=\sigma,\overrightarrow{\pi},\alpha}%
\tfrac{1}{2}\int_{k}[\ln G_{i}^{-1}(k)+D_{i}^{-1}(k;\phi,\alpha_{0}%
)G_{i}(k)-1]+V_{2}\text{ },%
\]
where $U(\phi,\alpha_{0})$ is the tree-level potential, $D_{i}(k;\phi
,\alpha_{0})$ are the tree-level propagators, $G_{i}(k)$ are the full
propagators. $V_{2}$ contains all two-particle-irreducible self-interaction terms,
but at one-loop level one has $V_{2}=0$.
Employing the stationary
conditions for the effective potential $\delta V_\mathrm{eff}/\delta\phi=\ \delta
V_\mathrm{eff}/\delta\alpha_{0}=$ $\delta V_\mathrm{eff}/\delta G_{i}(k)=0$
($i=\sigma,~\vec{\pi},~\alpha$), one gets the propagators $G_{i}(k)=-k^{2}+M_{i}%
^{2}\ =D_{i}^{-1}(k;\phi,\alpha_{0})$ with $M_{i}^{2}=m_{i}^{2}$ and the gap
equations
\begin{equation}
h=i\alpha_{0}\phi+\dfrac{4\phi}{N\varepsilon}\int_{k}G_{\sigma}(k)\text{ ,
}i\alpha_{0}=\dfrac{2}{N\varepsilon}\left[  \phi^{2}-\frac{N}{g^{2}}+\int
_{k}G_{\sigma}(k)+(N-1)\int_{k}G_{\pi}(k)\right]  \text{ .} \label{a1}%
\end{equation}
Eliminating $i\alpha_{0}$ by the previous expressions one finds the following
equations for the condensate and the masses:
\begin{align}
0  &  =\phi\left[  M_{\pi}^{2}+\dfrac{4}{N\varepsilon}\int_{k}G_{\sigma
}(k)\right]~~,~~~M_{\sigma}^{2}=M_{\pi}^{2}+\dfrac{4\phi^{2}%
}{N\varepsilon}\text{ },\nonumber\\
M_{\pi}^{2}  &  =\frac{2}{N\varepsilon}\left[  \phi^{2}-\frac{N}{g^{2}}%
+\int_{k}G_{\sigma}(k)+(N-1)\int_{k}G_{\pi}(k)\right]  \text{ }.
\end{align}
In the limit $\varepsilon\rightarrow0^{+}$: $\lim_{\varepsilon\rightarrow
0^{+}}M_{\sigma}^{2}=\infty$ $\rightarrow$ $\lim_{\varepsilon\rightarrow0^{+}%
}\int_{k}G_{\sigma}(k)/\varepsilon=0$. Thus, the gap equations read%
\begin{equation}
0=\phi M_{\pi}^{2}~~,~~~M_{\sigma}^{2}=M_{\pi}^{2}+\dfrac{4\phi^{2}%
}{N\varepsilon}~~,~~~\phi^{2}=\frac{N}{g^{2}}-(N-1)\int_{k}G_{\pi
}(k)\text{ .}%
\end{equation}
To satisfy the previous equation there are two possibilities:\newline

(i) $M_{\pi}^{2}=0,$ $\phi\neq0$ :%
\begin{equation}
\phi^{2}=\frac{N}{g^{2}}+(N-1)\int_{k}G_{\pi}(k)~,~~\phi^{2}-\frac
{N}{g^{2}}=(N-1)\int_{-\infty}^{\infty}\dfrac{dk}{2\pi}\dfrac{1}{\sqrt{k^{2}}%
}\left[  \exp\left(  \frac{\sqrt{k^{2}}}{T}\right)  -1\right]  ^{-1}\text{
.}\nonumber
\end{equation}
\newline

(ii) $\phi=0,$ $M_{\pi}^{2}\neq0$ :
\begin{equation}
M_{\sigma}^{2}=M_{\pi}^{2}=M^{2}~~,~~~\frac{N}{g^{2}}=N\int_{-\infty
}^{\infty}\dfrac{dk}{2\pi}\dfrac{1}{\sqrt{k^{2}+M^{2}}}\left[  \exp\left(
\frac{\sqrt{k^{2}+M^{2}}}{T}\right)  -1\right]  ^{-1}\text{ .}\nonumber
\end{equation}
There is no solution to the gap equations of case (i), since the integral
  $$\int\frac{[\exp[(\sqrt{k^{2}}/T)-1]^{-1}}{2\pi k} d k$$
is divergent, whereas
$\phi^{2}-N/g^{2}$ is finite. Therefore, contrary to the four-dimensional case
\cite{seel}, case (ii) is the right choice. This means that in two dimensions
there is no spontaneous symmetry breaking of the global $O(N)$ symmetry of the
nonlinear $O(N)$ model. This reflects the Mermin-Wagner-Coleman theorem, see
ref.~\cite{mermin}, which forbids spontaneous breakdown of a continuous
symmetry in a homogeneous system in one spatial dimension.

The model needs to be regularized and then renormalized. Here we just report
the results of these steps and we refer to ref.~\cite{our} for a detailed
treatment. The renormalized coupling constant $g_\mathrm{ren}$ develops a dependence
on the renormalization scale $\mu$:
\[
\mu\frac{dg_\mathrm{ren}^{2}}{d\mu}=-\frac{g_\mathrm{ren}^{4}}{2\pi}<0\;\;,
\]
thus showing that the theory is asymptotically free. At nonzero $T$ the mass
of the excitations is $M(T).$ For $T=0$ we obtain
\begin{equation}
M^{2}(T=0)=m^{2}=\mu^{2}\exp\left(  -\frac{4\pi}{g_\mathrm{ren}^{2}}\right)\;\;,
\end{equation}
which shows that a mass gap is generated and that the dilatation symmetry is
broken by quantum fluctuations (the energy scale $m$ is dynamically
generated). Then the function $M(T)$ rises linearly with $T$ when $T$
increases: this is an expected properties for a gas of quasiparticles and
takes place also in the deconfined phase of Yang-Mills theories, 
e.g.~ref.~\cite{quasiparticle} and refs.~therein.%

%TCIMACRO{\FRAME{ftbpFU}{3.8623in}{2.7086in}{0pt}{\Qcb{Comparison of the
%analytic expression for the pressure (the solid line is the one-loop result,
%the dashed line is the two-loop result) with lattice Monte-Carlo results
%obtained with the integral method.}}{}{o3pressure.eps}%
%width 3.8623in;  height 2.7086in;  depth 0pt;  original-width 5.0004in;
%original-height 3.4938in;  cropleft "0";  croptop "1";  cropright "1";
%cropbottom "0";  filename 'o3pressure.eps';file-properties "XNPEU";}}}%
%BeginExpansion
\begin{center}
\begin{figure}
\includegraphics[width=0.66\textwidth]{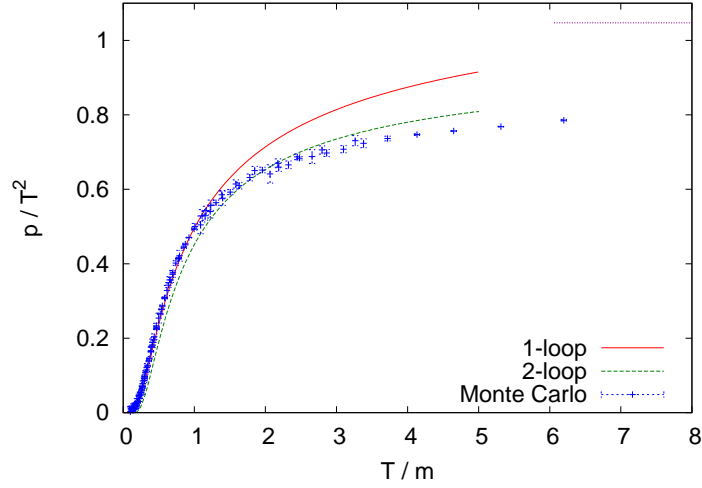}
\caption{Comparison of the analytic expression for the pressure (the solid
line is the one-loop result, the dashed line is the two-loop result) with
lattice Monte-Carlo results obtained with the integral method for the case $N=3$.
The upper right line marks the Stefan-Boltzmann limit $\pi/3$.}
\label{fig:pressure}
\end{figure}
\end{center}
%EndExpansion

We now turn to the pressure of the system, which is, up to a sign, identical
to the minimum of the effective potential: $p=-V_\mathrm{eff}^\mathrm{min}$. Its
renormalized form reads%
\begin{equation}
p(T)=N\dfrac{M(T)^{2}}{2g_{ren}^{2}}+N\int_{0}^{\infty}\dfrac{dk}{\pi
}\dfrac{k^{2}}{\omega_{k}}\dfrac{1}{\exp\left\{  \omega_{k}/T\right\}
-1}-N\frac{M(T)^{2}}{8\pi}\left(  1+\ln\frac{\mu^{2}}{M^{2}}\right)
+N\frac{m^{2}}{8\pi}\text{ .} \label{pressure}%
\end{equation}
Once the pressure is known one can compute the energy density $\rho$ and the
trace anomaly $\theta$ using the first principle of thermodynamics:
$\rho=Tdp/dT$ , $\theta=\rho-p$ .

The one-loop pressure of Eq.~(\ref{pressure}) is reported in\ Fig.~\ref{fig:pressure} and is
compared with a lattice study of this system \cite{our} for the case $N=3.$
One can notice that the result is good for small temperatures, but it does not
match the lattice data when the temperature increases. This result is indeed
expected: the one-loop expression in Eq.~(\ref{pressure}) reduces to
$T^{2}N\pi/6$ for large $T$, which corresponds to a free gas of $N$
non-interacting particles. This result cannot be correct, because the
constraint expressed by the $\delta$ function in Eq.~(\ref{z}) eliminates one
degree of freedom.

One can go beyond the one-loop results and study the pressure at the two-loop
level using the formalism developed in \cite{warringa}, see also ref.~\cite{our}.
A better agreement is obtained for large $T,$ although the low-$T$
domain is slightly worsened. Indeed, it is possible to show analytically that
for large $T$ the two-loop pressure approaches the correct value
$T^{2}(N-1)\pi/6$.

\section{Conclusions}

In this work we have studied the $O(N)$ model in 1+1 dimensions at nonzero
$T.$ We have described how a mass gap is generated for $T=0$, thus demonstrating 
the occurrence of dimensional transmutation. We have then shown
explicitly the analytic results for the pressure at the one-loop level and
briefly commented on the results at the two-loop level. Finally, these
analytic expressions for the pressure have been compared with lattice results,
see Fig.~\ref{fig:pressure}.

More advanced analytical calculations can be performed in the
future and compared to lattice data: in fact, this system represents a good
tool to test nonperturbative approaches, which can then be applied to the more
difficult case of Yang-Mills theories in four dimensions.


\begin{thebibliography}{99}

%%%%%%%%%%%%%%%%%%%%%%%%%%%%%%%%%%%%%%%%%%%%%%%
%\cite{Novikov:1984ac}
\bibitem[Novikov(1984)]{novikov}V.~A.~Novikov, M.~A.~Shifman, A.~I.~Vainshtein
and V.~I.~Zakharov,
%``Two-Dimensional Sigma Models: Modeling Nonperturbative Effects of Quantum Chromodynamics,''
Phys.\ Rept.\ \textbf{116} (1984) 103 [Sov.\ J.\ Part.\ Nucl.\ \textbf{17}
(1986) 204]; Fiz.\ Elem.\ Chast.\ Atom.\ Yadra \textbf{17} (1986) 472.
%%%%%%%%%%%%%%%%%%%%%%%%%%%%%%%%%%%%%%%%%%%%%%%%%%

%\cite{Wolff:1989hv}
\bibitem[Wolff(1990)]{wolff}U.~Wolff,
%``Asymptotic Freedom And Mass Generation In The O(3) Nonlinear Sigma Model,''
Nucl.\ Phys.\ B \textbf{334} (1990) 581.
%%CITATION = NUPHA,B334,581;%%
%%%%%%%%%%%%%%%%%%%%%%%%%%%%%%%%%%%

%%CITATION = PRPLC,116,103;%%%%%%%%%%%%%%%%%%%%%%%%%%%%%%%%
%\cite{Bruckmann:2007zh}
\bibitem[Bruckmann(2008)]{bruckmann}F.~Bruckmann,
%``Instanton constituents in the O(3) model at finite temperature,''
Phys.\ Rev.\ Lett.\ \textbf{100} (2008) 051602 [arXiv:0707.0775 [hep-th]];
%%CITATION = ARXIV:0707.0775;%%
%%%%%%%%%%%%%%%%%%%%%%%%%%%%%%%
J.~O.~Andersen, D.~Boer and H.~J.~Warringa,
%``The Effects of quantum instantons on the thermodynamics of the CP**(N-1) model,''
Phys.\ Rev.\ D \textbf{74} (2006) 045028 [hep-th/0602082];
%%CITATION = HEP-TH/0602082;%%
%%%%%%%%%%%%%%%%%%%%%%%%%%
%\cite{Affleck:1980mb}
I.~Affleck,
%``Testing The Instanton Method,''
Phys.\ Lett.\ B \textbf{92} (1980) 149.
%%CITATION = PHLTA,B92,149;%%
%%%%%%%%%%%%%%%%%%%%%%%%%%%


%\cite{Hofmann:2005dt}
\bibitem[Hofmann(1995)]{ralf1}R.~Hofmann,
%``Nonperturbative approach to Yang-Mills thermodynamics,''
Int.\ J.\ Mod.\ Phys.\ A \textbf{20} (2005) 4123 [hep-th/0504064];
%%CITATION = HEP-TH/0504064;%%
%\cite{Herbst:2004nk}
%\bibitem{Herbst:2004nk}
U.~Herbst and R.~Hofmann,
%``Asymptotic freedom and compositeness,''
ISRN High Energy Phys.\ \textbf{2012} (2012) 373121 [hep-th/0411214];
%%CITATION = HEP-TH/0411214;%%
%\cite{Giacosa:2007zz}
%\bibitem{Giacosa:2007zz}
F.~Giacosa and R.~Hofmann,
%``Thermal ground state in deconfining Yang-Mills thermodynamics,''
Prog.\ Theor.\ Phys.\ \textbf{118} (2007) 759 [hep-th/0609172].
%%CITATION = HEP-TH/0609172;%%
%%%%%%%%%%%%%%%%%%%%%%%%%%%%%%%%%%%%%%%%%%%%%%%%%%


\bibitem[Pisarski(1981)]{pisarski}D.~J.~Gross, R.~D.~Pisarski and L.~G.~Yaffe,
%``QCD and Instantons at Finite Temperature,''
Rev.\ Mod.\ Phys.\ \textbf{53} (1981) 43.
%%CITATION = RMPHA,53,43;%%
%%%%%%%%%%%%%%%%%%%%%%%%%%%%%%%%%%%%%%%%%%
%\cite{Andersen:2003va}


\bibitem[Cornwall(1974)]{cjt}J.~M.~Cornwall, R.~Jackiw, and E.~Tomboulis,
Phys.~Rev.~D \textbf{10} 2428 (1974).
%%%%%%%%%%%%%%%%%%%%%%%%%%%


\bibitem[Giacosa(2011)]{our} E.~Seel, D.~Smith, S.~Lottini and F.~Giacosa,
 %``Thermodynamics of the O(3) model in 1+1 dimensions: lattice vs. analytical results,''
  arXiv:1209.4243 [hep-ph].
  %%CITATION = ARXIV:1209.4243;%%


%\cite{Seel:2011ju}
\bibitem[Seel(2012)]{seel}E.~Seel, S.~Struber, F.~Giacosa and
D.~H.~Rischke,
%``Study of chiral symmetry restoration in linear and nonlinear O(N) models using the auxiliary field method,''
arXiv:1108.1918 [hep-ph].
%%CITATION = ARXIV:1108.1918;%%
%%%%%%%%%%%%%%%%%%%%%%%%%%%%%%%%%%%%%%%%%%%%%%%%%%%%%%%%%%
%\cite{Giacosa:2010vz}

%\cite{Mermin:1966fe}
\bibitem[Mermin(1966)]{mermin}N.~D.~Mermin and H.~Wagner,
%``Absence of ferromagnetism or antiferromagnetism in one-dimensional or two-dimensional isotropic Heisenberg models,''
Phys.\ Rev.\ Lett.\ \textbf{17} (1966) 1133;
%%CITATION = PRLTA,17,1133;%%
%\cite{Coleman:1973ci}
%\bibitem{Coleman:1973ci}
S.~R.~Coleman,
%``There are no Goldstone bosons in two-dimensions,''
Commun.\ Math.\ Phys.\ \textbf{31} (1973) 259.
%%CITATION = CMPHA,31,259;%%
%%%%%%%%%%%%%%%%%%%%%%%%%%%%%%%%%%


\bibitem[Giacosa(2011)]{quasiparticle}F.~Giacosa,
%``Analytical study of a gas of gluonic quasiparticles at high temperature: Effective mass, pressure and trace anomaly,''
Phys.\ Rev.\ D \textbf{83} (2011) 114002 [arXiv:1009.4588 [hep-ph]];
%%CITATION = ARXIV:1009.4588;%%
%\cite{Brau:2009mp}
%\bibitem{Brau:2009mp}
F.~Brau and F.~Buisseret,
%``Glueballs and statistical mechanics of the gluon plasma,''
Phys.\ Rev.\ D \textbf{79} (2009) 114007 [arXiv:0902.4836 [hep-ph]];
%%CITATION = ARXIV:0902.4836;%%
%\cite{Peshier:1995ty}
%\bibitem{Peshier:1995ty}
A.~Peshier, B.~Kampfer, O.~P.~Pavlenko and G.~Soff,
%``A Massive quasiparticle model of the SU(3) gluon plasma,''
Phys.\ Rev.\ D \textbf{54} (1996) 2399;
%%CITATION = PHRVA,D54,2399;%%
%\cite{Castorina:2007qv}
%\bibitem{Castorina:2007qv}
P.~Castorina and M.~Mannarelli,
%``Effective degrees of freedom and gluon condensation in the high temperature deconfined phase,''
Phys.\ Rev.\ C \textbf{75} (2007) 054901 [hep-ph/0701206 [hep-ph]].
%%CITATION = HEP-PH/0701206;%%
%%%%%%%%%%%%%%%%%%%%%%%%%%%%%%%%%%%%%%%%%%
%\cite{Gross:1980br}


\bibitem[Warringa(2004)]{warringa}J.~O.~Andersen, D.~Boer and H.~J.~Warringa,
%``Thermodynamics of the O(N) nonlinear sigma model in (1+1)-dimensions,''
Phys.\ Rev.\ D \textbf{69} (2004) 076006 [hep-ph/0309091].
%%CITATION = HEP-PH/0309091;%%
%%%%%%%%%%%%%%%%%%%%%%%%%%%%%%%%
%\cite{Andersen:2004ae}

\end{thebibliography}
\end{document}